%% file: main.tex
\documentclass[sigconf]{acmart}
\usepackage{booktabs} 

\usepackage[english]{babel}
\usepackage{moresize}
\usepackage{amsmath}
\usepackage{algorithmic}
\usepackage{balance}
\usepackage{comment}
\usepackage{paralist}
\usepackage{bm}
\usepackage{pgfplots}
\usetikzlibrary{pgfplots.dateplot}

\usepackage{flushend}
\usepackage[english]{babel}
\usepackage[latin1]{inputenc}
\usepackage{mathrsfs}
\usepackage{graphicx}

\usepackage{amssymb}
\usepackage{amsfonts}
\usepackage{url}
\usepackage{longtable}
\usepackage{rotating}
\usepackage{multirow}
\usepackage{mathrsfs}
\usepackage{subfigure}
\usepackage{enumitem}
\usepackage[linesnumbered,algoruled,boxed,lined]{algorithm2e}
\usepackage{adjustbox}
\usepackage{hyperref}
\usepackage{pgfplots}
\usetikzlibrary{pgfplots.dateplot}
\usepackage{filecontents}
\usepackage{subcaption}

\definecolor{tblue}{RGB}{31,119,180}
\definecolor{torange}{RGB}{255,127,14}
\definecolor{tgreen}{RGB}{44,160,44}
\definecolor{tred}{RGB}{214,39,40}
\definecolor{tpurple}{RGB}{148,103,189}
\newcommand{\hide}[1]{} 

\newcommand{\ie}{\textit{i}.\textit{e}.}
\newcommand{\eg}{\textit{e}.\textit{g}.} 
\newcommand{\wrt}{\textit{w}.\textit{r}.\textit{t}} 

\newcommand{\e}{{\mathbf{e}}}

\def\model{SelfGNN}

\copyrightyear{2024}
\acmYear{2024}
\setcopyright{acmlicensed}\acmConference[SIGIR'24]{Proceedings of the 47th International ACM SIGIR Conference on Research and Development in Information Retrieval}{July 14--18, 2024}{Washington, DC, USA}
\acmBooktitle{Proceedings of the 47th International ACM SIGIR Conference on Research and Development in Information Retrieval (SIGIR'24), July 14--18, 2024, Washington, DC, USA}
\acmDOI{10.1145/3626772.3657716}
\acmISBN{979-8-4007-0431-4/24/07}


\begin{document}

\title{SelfGNN: Self-Supervised Graph Neural Networks for \\ Sequential Recommendation}

\author{Yuxi Liu}
\orcid{0009-0006-0430-373X}
\affiliation{%
  \institution{School of Electronic and Computer Engineering, Peking University}
  \city{Shenzhen}
  \country{China}
}
\email{liuyuxi.tongji@gmail.com}

\author{Lianghao Xia}
\affiliation{%
  \institution{Computer Science Department, University of Hong Kong}
  \city{Hong Kong}
  \country{China}}
\email{aka\_xia@foxmail.com}

\author{Chao Huang}
\authornote{Chao Huang is the Corresponding Author.}
\affiliation{%
  \institution{Computer Science Department \& IDS, University of Hong Kong}
  \city{Hong Kong}
  \country{China}}
\email{chaohuang75@gmail.com}

\begin{abstract}
Sequential recommendation effectively addresses information overload by modeling users' temporal and sequential interaction patterns. To overcome the limitations of supervision signals, recent approaches have adopted self-supervised learning techniques in recommender systems. However, there are still two critical challenges that remain unsolved. Firstly, existing sequential models primarily focus on long-term modeling of individual interaction sequences, overlooking the valuable short-term collaborative relationships among the behaviors of different users. Secondly, real-world data often contain noise, particularly in users' short-term behaviors, which can arise from temporary intents or misclicks. Such noise negatively impacts the accuracy of both graph and sequence models, further complicating the modeling process. To address these challenges, we propose a novel framework called Self-Supervised Graph Neural Network (\model) for sequential recommendation. The \model\ framework encodes short-term graphs based on time intervals and utilizes Graph Neural Networks (GNNs) to learn short-term collaborative relationships. It captures long-term user and item representations at multiple granularity levels through interval fusion and dynamic behavior modeling. Importantly, our personalized self-augmented learning structure enhances model robustness by mitigating noise in short-term graphs based on long-term user interests and personal stability. Extensive experiments conducted on four real-world datasets demonstrate that \model\ outperforms various state-of-the-art baselines. Our model implementation codes are available at \color{blue}\url{https://github.com/HKUDS/SelfGNN.}
\end{abstract}




\maketitle

\input{intro}
\input{model}
\input{solution}
\input{eval}
\input{conclusion}

\clearpage

\bibliographystyle{ACM-Reference-Format}
\balance
\bibliography{sample-base}


\end{document}

%% file: intro.tex
\section{Introduction}
\label{sec:intro}
Recommender systems have emerged as a crucial field for addressing the information overload issue, providing benefits to both users and service providers such as online shopping platforms (\eg, Tmall, Amazon) \cite{ge2020understanding}, video websites (\eg, TikTok, YouTube)~\cite{zhang2021mining,wei2023multi}, and social networks (\eg, Gowalla, Facebook) \cite{peng_2020,User7470266}. A wide range of solutions has been proposed, such as factorization-based methods \cite{Salakhutdinov_2007,Heterogeneous8355676} and neural collaborative filtering models \cite{NCF,NAIS}. Recent research has introduced Graph Neural Networks (GNNs) to capture high-order relations over user-item interaction graphs for encoding user preferences. For instance, NGCF \cite{NGCF} utilizes Graph Convolutional Networks (GCNs) to model user-item interactions, and LightGCN \cite{LightGCN} simplifies and enhances the power of GCNs. Recent efforts have also been dedicated to dynamic GNNs in the context of recommendation systems. For instance, DGCF \cite{li2020dynamic} introduces dynamic graphs into recommendation, while DGSR \cite{zhang2022dynamic} connects different user sequences through dynamic graph structures.

\begin{figure}[t]
    \centering
    \includegraphics[width=0.46\textwidth]{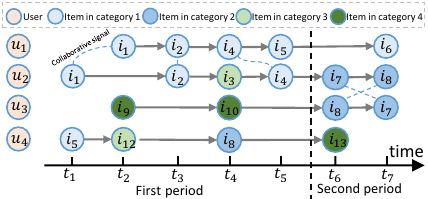}
    \vspace{-0.10in}
    \caption{Illustrative examples reflecting collaborative relations (between $u_1$ and $u_2$ in first period, between $u_2$ and $u_3$ in second period), temporal intents ($u_2$ adopting $i_3$), interest shift ($u_3$), and user with variable interests ($u_4$).}
    \label{fig:example}
    \vspace{-0.2in}
\end{figure}

One important line of research in recommender systems is sequential recommendation, which involves analyzing users' temporal interaction patterns and predicting their future behaviors. Various deep learning methods have been developed to address this task. For example, GRU4Rec \cite{GRU4Rec} utilizes recurrent neural networks (RNN) to model sequential patterns. Similarly, SASRec \cite{SASRec} and Bert4Rec \cite{Bert4Rec} leverage self-attention mechanisms to capture item-item pairwise correlations. To address the data scarcity challenge, 

Despite the significant progress made in graph and sequential recommenders with SSL, there are still several challenges that remain unaddressed (illustrated in Fig~\ref{fig:example}). Firstly, existing long-short-term sequential recommenders~\cite{SURGE, CLSR} often mainly focus on encoding single user sequence, which overlook the essential periodical collaborative relationships between users~\cite{Fan_2021}. Traditional GNNs solely rely on static data. Although some work has constructed dynamic and temporal GNNs, they overlook the consideration of both long and short-term aspects in GNNs, nor do they address the construction of a more robust dynamic learning paradigm.

Secondly, the impressive results achieved by existing SSL methods heavily depend on high-quality data~\cite{xia2023automated}. However, real-world data often contains noise, such as users' misclicks, temporary intents, or interest shifts~\cite{Yuta_2019}. Existing models tend to amplify such noises through multi-hop graph message propagation or multi-layer pairwise self-attention, which can greatly impact the representation of other nodes~\cite{Sun_2021}. It is worth noting that the notion of "noise" varies for different users, as some users have diverse interests while others have more stable preferences. Current denoising approaches lack the ability to effectively handle user-specific noise and often rely on random data augmentation. These challenges hinder the accurate representation of user behaviors. \\ \vspace{-0.12in}

\noindent \textbf{Contribution.} In light of the aforementioned challenges, this work proposes a \underline{Self}-Supervised \underline{G}raph \underline{N}eural \underline{N}etwork (\model) for sequential recommendation. \model\ aims to capture user interests by incorporating dynamic collaborative information and personalizing short-term denoising through self-augmented learning. \model\ is built upon three key paradigms. (1) \textbf{Short-term Collaborative Graphs Encoding.} The global user-item graph is divided into several short-term graphs based on time intervals. GCNs are employed to propagate collaborative high-order information. For instance, in Figure \ref{fig:example}, $u_1$ and $u_2$ have significant collaborative signals in the first period, while $u_3$ has similar behaviors with $u_2$ in the second period. In this way, we not only capture the collaborative patterns but also include short-term temporal information.
(2) \textbf{Multi-level Long-term Sequential Learning.} We conduct dual-level sequence modeling, which forms the interest with different granularity complement each other. At the time interval level, the features learned from different short-term graphs are treated as sequences to generate long-term collaborative representations for users. At the instance level, the complete sequence of user behavior is modeled by self-attention mechanisms. (3) \textbf{Personalized Self-Augmented Learning for Denoising.} A self-augmented learning task is designed to correct the corresponding relationships in the short-term graphs based on the long-term user interests. This task learns personalized weights for users and adjusts the noise level according to users' interest stability (\eg, $u_2$ with a stable tendency and $u_4$ with variable interests in Figure \ref{fig:example}). Consequently, the \model\ can discriminate abnormal interactions in the short term and mitigate their impact by adapting to different users. 

The key contributions can be summarized as follows: 
\begin{itemize}[leftmargin=*]

\item We introduce a novel self-supervised graph recommender framework, which effectively captures dynamic user interests by integrating interval-level periodical collaborative relationship learning and attentive instance-level sequential modeling.

\item Our method incorporates a personalized self-augmented learning component, enabling effective denoising of noisy interactions in sparse short-term graphs. It also generates personalized weights to adapt to different users, enhancing the model's ability to cater to individual preferences and stability.

\item Our extensive experiments on real-world datasets demonstrate that \model\ outperforms various baseline models. Furthermore, the robust performance of our framework is thoroughly validated, further underscoring its effectiveness and practical utility.

\end{itemize}

%% file: model.tex
\section{Related Work}
\label{sec:model}
\subsection{GNNs for Recommendation}
Recent studies have introduced diverse Graph Neural Network architectures for encoding user-item interactions within graph-structured data~\cite{rao2015collaborative,tang2024higpt}. Networks like GC-MC \cite{berg2017graph} and NGCF \cite{NGCF} are adept at capturing high-order collaborative signals through extensive information propagation. Innovations such as LightGCN \cite{LightGCN} and GCCF \cite{chen2020revisiting} have streamlined and enhanced the foundational GCN architecture. These have also inspired temporal GNN methods for sequential recommendation, exemplified by SRGNN \cite{SRGNN}, GCE-GNN \cite{GCE-GNN}, and TGSREC \cite{fan2021continuous}, each contributing novel graph construction and attention mechanisms. Dynamic graph design has been another area of focus, with models like DGCF \cite{li2020dynamic} and TG-MC \cite{bonet2021temporal} integrating temporal dynamics into user and item representation learning. DGSR \cite{zhang2022dynamic} and SURGE \cite{SURGE} further advance the field by explicitly modeling dynamic user-item interaction graphs.

However, existing methods have yet to fully address the integration of long and short-term user features within a robust sequential GNN model. Our proposed \model\ tackles this challenge by revisiting GNN with a self-augmented learning paradigm. Our proposed new \model, grounded in long-term feature supervision, aims to reduce noise in short-term graphs and fortify the model's ability to learn from both immediate and enduring user preferences.

\vspace{-0.12in}
\subsection{Sequential Recommender Systems}
User interaction sequences play a pivotal role in recommender systems, as they highlight the chronological patterns of user behavior. These patterns provide critical insights into both the short-term dependencies of user interactions and the gradual shifts in user preferences. As a result, a multitude of studies have been devoted to uncovering the dynamics of user preferences over time, with sequential recommendation being a notable area, such as those developed using Recurrent Neural Networks like GRU4Rec \cite{GRU4Rec}. In recent years, the transformative capabilities of the Transformer architecture have catalyzed the broad application of self-attention mechanisms. These mechanisms excel at capturing the intricate correlations between items within a user's sequence. Notable implementations include SASRec \cite{SASRec}, Bert4Rec \cite{Bert4Rec}, TiSASRec \cite{TiSASRec} and MBHT~\cite{yang2022multi}. Moreover, CLSR \cite{CLSR}, while adept at discerning both short- and long-term user interests within sequential recommendations, falls short in thoroughly tackling the challenge of data noise. This aspect remains a significant area for improvement, as addressing it could lead to more accurate recommendation results.

\subsection{Recommender Systems with SSL}
Self-supervised learning has gained significant traction in bolstering the capabilities of recommendation systems, evident in both graph-based and sequential models~\cite{jing2023contrastive,sslrec2024}. Its popularity stems from its capacity to produce auxiliary supervision signals that effectively tackle the prevalent challenge of data sparsity~\cite{Liu_2021, Self-Supervised10144391,jiang2023adaptive}. For instance, SGL \cite{SGL} capitalizes on node self-discrimination through random augmentations of graph structures. CoTRec \cite{CoTRec} employs a dual-level contrastive learning strategy across two similar item relation graphs. Meanwhile, AutoCF \cite{xia2023automated} designs autonomous masked autoencoding framework to offer robust self-supervised signals. In the realm of sequential recommendations, ICLRec \cite{ICLRec} and CoSeRec \cite{CoSeRec} enhance interaction sequences by employing techniques such as cropping, masking, or reversing, followed by the application of contrastive learning across these diversified views. Distinct from these methodologies, our self-supervised learning framework is uniquely tailored to filter out noise from short-term interactions. This is achieved by embedding a deep understanding of long-term user behavior into the model, thereby refining the accuracy and relevance of the recommender system's output. 


%% file: solution.tex
\section{Methodology}
\label{sec:solution}

\begin{figure*}[t]
\centering
\includegraphics[width=\textwidth]{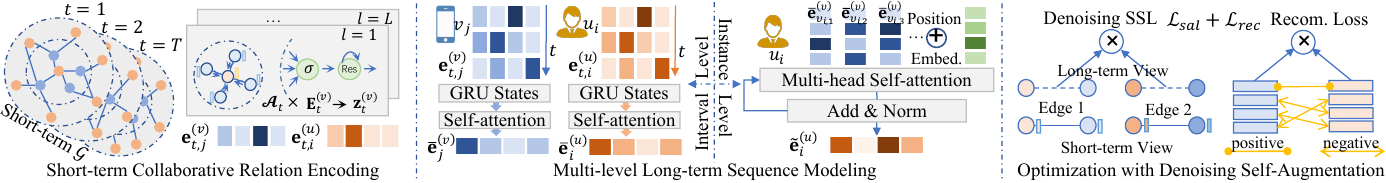}
\vspace{-0.25in}
\caption{Overall framework of the proposed \model\ model.} 
\label{fig:overall}
\vspace{-0.15in}
\end{figure*}
In this section, we introduce our proposed \model\ framework and illustrate the overall model architecture in Figure \ref{fig:overall}.
\subsection{Notations and Formalization}
In this paper, we propose to combine the strength of GNNs in capturing high-order collaborative relations and the strength of sequence modeling for capturing users' preference shifts, with the help of self-supervised learning techniques. In such scenario, we define $\mathcal{U}=\{u_1,...,u_i,...,u_I\}(|\mathcal{U}|=I)$ and $\mathcal{V}=\{v_1,...,v_j,...,v_J\}$ $(|\mathcal{V}|=J)$ to represent the set of users and items respectively. 

\noindent \textbf{The method of segmenting user interaction data into different periods.} We partition the entire time interval in the dataset according to timestamp, ranging from the earliest occurrence at time $t_b$ to the latest occurrence at time $t_e$, into an average of $T$ short-term time intervals, where $T$ is a tunable hyperparameter. The duration of each time interval is $(t_e - t_b) / T$. The adjacent matrix $\mathcal{A}_t\in \mathbb{R}^{I\times J}$ indicates the implicit relationships between each user and the items they interacted with in the ($t$)-th period. Each entry $\mathcal{A}_{t,i,j}$ in $\mathcal{A}_{t}$ is set as 1 if user $u_{i}$ has adopted item $v_j$, and $\mathcal{A}_{t,i,j}=0$ otherwise.

\noindent\textbf{Formalization}. Given partitioned graph-structured interactions $\{\mathcal{A}_t|1\leq t\leq T\}$, predict the future interactions $\mathcal{A}_{T+1}$, with the following abstracted optimization objective:
\begin{align}
    &\mathop{\arg\min}\nolimits_{\mathbf{\mathbf{\Theta}}_f, \mathbf{\mathbf{\Theta}}_g} ~\mathcal{L}_{rec}(\hat{\mathcal{A}}_{T+1}, \mathcal{A}_{T+1}) + \mathcal{L}_{sal}(\textbf{E}_{s}, \textbf{E}_{l})\nonumber\\
    &\hat{\mathcal{A}}_{T+1} = f(\textbf{E}_{l}, \textbf{E}_{s};\mathbf{\Theta}_f),~~~ \textbf{E}_{l}, \textbf{E}_{s} = g(\{\mathcal{A}_t\}; \mathbf{\Theta}_g)
\end{align}
where $\mathcal{L}_{rec}$ denotes the loss function measuring the error between predictions $\hat{\mathcal{A}}_{T+1}$ and ground-truth ${\mathcal{A}}_{T+1}$, and $\mathcal{L}_{sal}$ denotes the SSL objective for aligning the short-term hidden embeddings $\textbf{E}_{s}$ and the long-term embeddings $\textbf{E}_{l}$. $\hat{\mathcal{A}}_{T+1}$ is acquired by the prediction function $f$ with parameter set $\mathbf{\Theta}_f$, based on the short and long-term information $\textbf{E}_{s}, \textbf{E}_{l}$ encoded by the encoder $g$ with parameters $\mathbf{\Theta}_g$.

\subsection{Short-term Collaborative Relation Encoding}
Firstly, we perform collaborative information learning for each short-term interaction graph. Specifically, we project each user $u_i$ and item $v_j$ in each time slot $t$, into a $d$-dimensional latent space with $\textbf{e}_{t,i}^{(u)}, \textbf{e}_{t,j}^{(v)}\in\mathbb{R}^d$. These embedding vectors compose embedding matrices $\textbf{E}_t^{(u)}\in\mathbb{R}^{I\times d}$ and $\textbf{E}_t^{(v)}\in\mathbb{R}^{J\times d}$ for the $t$-th period, respectively. Inspired by LightGCN~\cite{LightGCN}, we employ the following simplified GCN for short-term graph modeling:
\begin{equation}
    {\mathbf{z}_{t,i}^{(u)}}=\sigma( {\mathcal{A}_t}_{i,*}\cdot \mathbf{E}_t^{(v)}),~~~ {\mathbf{z}_{t,j}^{(v)}}=\sigma( {\mathcal{A}_t}_{j,*}\cdot \mathbf{E}_t^{(u)})
\end{equation}
where ${\mathbf{z}_{t,i}^{(u)}},{\mathbf{z}_{t,j}^{(v)}}\in \mathbb{R}^{d}$ denote the information aggregated from neighboring nodes to the centric node $u_i$ and $v_j$ and $\sigma(\cdot)$ denotes the LeakyReLU function. We adopt random drops to the edges of the graphs to mitigate the overfitting problem. We stack such GCN layers for high-order propagation. 

In the $l$-th layer, the message-passing process is: 
\begin{equation}
    \mathbf{e}^{(u)}_{t, i,l}=\mathbf{z}^{(u)}_{t,i,l}+\mathbf{e}^{(u)}_{t, i,l-1},~~~\mathbf{e}^{(v)}_{t,j,l}=\mathbf{z}^{(v)}_{t, j,l}+\mathbf{e}^{(v)}_{t, j,l-1}
\end{equation}
where $\mathbf{e}^{(u)}_{t, i,l}, \mathbf{e}^{(v)}_{t,j,l}\in\mathbb{R}^d$ represent the embeddings of $u_i$ and $v_j$ at the $l$-th GNN layer for the $t$-th time period.
Since different layers emphasize different propagation ranges, we concatenate them to obtain the final short-term embedding:
\begin{equation}
    \mathbf{e}_{t,i}^{(u)}={\mathbf{e}_{t,i,1}^{(u)}}||\cdots||{\mathbf{e}_{t,i,L}^{(u)}}, ~~~
    {\mathbf{e}_{t,j}^{(v)}}={\mathbf{e}_{t,j,1}^{(v)}}||\cdots||{\mathbf{e}_{t, j,L}^{(v)}}
\end{equation}

\subsection{Multi-level Long-term Sequence Modeling}
\label{sec:interval-level}
In this section we aim to model the long-term relationship at two levels: i) Interval-Level Sequential Pattern Modeling, which integrates short-term features into long-term embeddings based on temporal attention, capturing the dynamic changes from period to period, and ii) Instance-Level Sequential Pattern Modeling, which learns pairwise relations between specific item instances directly.

\subsubsection{\bf Interval-Level Sequential Pattern Modeling}
For every user and item, we construct a chronological embedding sequence based on their short-term embeddings. To inject the temporal information into our interval-level sequence, we utilize the Gated Recurrent Unit (GRU) network~\cite{an-etal-2019-neural}, instead of the position encoding used in Transformer because position embedding would be too simplistic to capture the temporal information when dealing with a small number of intervals in our work. The sequences for user $u_i$ and item $v_j$ are defined as follows:
\begin{align}
    S_{i}^{intv} &= (\textbf{h}_{1,i}^{(u)}, \cdots, \textbf{h}_{t,i}^{(u)}, \cdots, \textbf{h}_{T,i}^{(u)})\nonumber\\
    S_{j}^{intv} &= (\textbf{h}_{1,j}^{(v)}, \cdots, \textbf{h}_{t,j}^{(v)}, \cdots, \textbf{h}_{T,j}^{(v)})\nonumber\\
    \textbf{h}^{(u)}_{t,i} = \text{GRU}&(\textbf{e}_{t,i}^{(u)}, \textbf{h}_{t-1,i}^{(u)}),~~ \textbf{h}^{(v)}_{t,j} = \text{GRU}(\textbf{e}_{t,j}^{(v)}, \textbf{h}_{t-1,j}^{(v)})
\end{align}
where $\textbf{h}_{t,i}^{(u)}, \textbf{h}_{t,j}^{(v)}\in\mathbb{R}^d$ denote the hidden states of GRU at the $t$-th time slot. Then \model\ devices the multi-head dot-product attention, denoted as $\text{Self-Att}(\cdot)$, over the interval-level sequences $S_i^{intv}$ and $S_j^{intv}$, respectively, to excavate the dynamic patterns: 
\begin{align}
    \bar{\textbf{H}}_i^{(u)} = \text{Self-Att}(S_i^{intv}),~~~
    \bar{\textbf{H}}_j^{(v)} = \text{Self-Att}(S_j^{intv})
\end{align}

Finally, we sum up the features to obtain the long-term features:
\begin{align}
    \bar{\textbf{e}}_i^{(u)} = \sum_{t=1}^T \bar{\textbf{H}}_{i,t}^{(u)},~~~~
    \bar{\textbf{e}}_j^{(v)} = \sum_{t=1}^T \bar{\textbf{H}}_{j,t}^{(v)}
\end{align}
where $\bar{\textbf{e}}_i^{(u)}, \bar{\textbf{e}}_j^{(v)}\in\mathbb{R}^d$ refer to the output embedding vectors for $u_i$ and $v_j$, containing their long-term patterns. Compared to only modeling instance sequences, our interval-level modeling enriches the module with periodical collaborative signals. 

\subsubsection{\bf Instance-Level Sequential Pattern Modeling}
Motivated by the success of the self-attention mechanism~\cite{SASRec, Bert4Rec}, we also augment our \model\ with the self-attention network directly over sequences containing users' interacted item instances. Denoting the $m$-th interacted item of user $u_i$ as $v_{i,m}$, where $m\in\{1, 2, ..., M\}$ and $M$ represents the maximum interaction length, we compose the following sequences for recording $u_i$'s actions:
\begin{align}
    S_{i,0}^{inst} = (\bar{\textbf{e}}_{v_{i,1}}^{(v)} + \textbf{p}_1, \cdots, 
    \bar{\textbf{e}}_{v_{i,M}}^{(v)} + \textbf{p}_M)
\end{align}
where $\bar{\textbf{e}}_{v_{i,m}}^{(v)} \in \mathbb{R}^d$ refers to the embedding for item $v_{i,m}$, and $\textbf{p}_m\in\mathbb{R}^d$ denotes the learnable positional embedding for the $m$-th position. We employ $L_{a}$ layers of self-attention with residual connections to capture the sequential patterns: 
\begin{align}
    S_{i,l}^{inst} = \sigma(\text{Self-Att}(S_{i,l-1}^{inst})) + S_{i,l-1}^{inst},~~~
    \tilde{\textbf{e}}_i^{(u)} = \sum S_{i,L_{a}}^{inst}
\end{align}
where $S_{i,l}^{inst}$ denotes sequence for $u_i$ in the $l$-th self-attention iteration. Based on the processed sequences, we sum the element embeddings and use the above $\tilde{\textbf{e}}_i^{(u)} \in \mathbb{R}^d$ to represent user $u_i$ with instance-level sequential correlations.

\subsection{Multi-view Aggregation and Prediction}
Before prediction, we aggregate the multi-views user features derived from the instance-level and the interval-level approach, and make final predictions as follows:
\begin{align}
    \hat{\mathcal{A}}_{T+1, i,j}={\hat{\e}_{i}^{(u)\top}} \cdot {\bar{\e}_{j}^{(v)}},~~~\hat{\e}^{(u)}_i={\bar{\e}_i^{(u)}}+{\tilde{\e}_{i}^{(u)}}
\end{align}
where $\hat{\mathcal{A}}_{T+1,i,j}\in\mathbb{R}$ denotes the prediction for $u_i$ interacting with $v_j$, in the future ($T+1$)-th time slot by calculating the similarity between user and item. Nextly, positive samples consist of items that users have interacted with, while negative samples consist of items not interacted with. Imposing a restriction to prevent the predicted values from becoming arbitrarily large, we optimize the following loss function as follows:
\begin{align}
    \mathcal{L}_{rec}= \sum_{i=1}^{I} \sum_{k=1}^{N_{pr}}{\rm{max}}(0,1-\hat{\mathcal{A}}_{T+1,i,p_k}+\hat{\mathcal{A}}_{T+1, i, n_k})
\end{align}
where $N_{pr}$ is the number of samples, $p_k,n_k$ represents the ($k$)-th positive and negative item index respectively.

\subsection{Personalized Denoising Self-Augmentation}
\begin{figure}[t]
    \centering
    \includegraphics[width=\columnwidth]{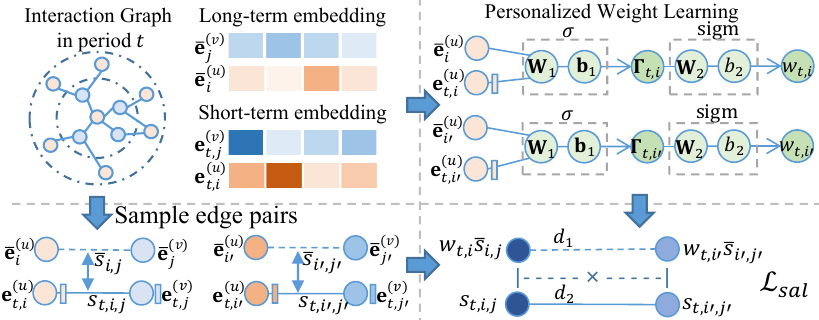}
    \vspace{-10pt}
    \caption{Workflow of the personalized SSL paradigm.} 
    \label{fig:ssl}
    \vspace{-10pt}
\end{figure}

To alleviate the ubiquitous data sparsity and data noise problem in users' sequential behavior data, our \model\ is further enhanced by a personalized denoising self-supervised learning task. The term "noise" refers to temporary intents or misclicks, which cannot be considered as long-term user interests or new recent points of interest for predictions. Specifically, our SSL task focuses on filtering short-term non-inherent user preferences using long-term behavior patterns. This design is based on the observation that, users' behaviors may be motivated by short-term random interests, \eg, a user who is not in favor of hiking may also buy hiking shoes and energy drinks because of a one-time activity. Such noisy behavior data may disturb the modeling of users' true interests in the long term. Furthermore, to accurately identify such noisy short-term behaviors, we personalize the denoising SSL task with different strengths for different users, which caters to users with different levels of interest variety, as shown in Figure~\ref{fig:ssl}.

In particular, for each training sample of our denoising SSL, we randomly sample two observed user-item edges $(u_i, v_j)$ and $(u_{i'}, v_{j'})$ from the short-term graphs $\mathcal{A}_{t}$, and align the pairwise likelihood difference scores given. Taking $(u_i, v_j)$ as an example, the formula is as follows:
\begin{align}
    s_{t,i,j} = \sum_{k=1}^d \sigma({e}_{t,i,k}^{(u)} \cdot {e}_{t,j,k}^{(v)}),~~~~
    \bar{s}_{i,j} = \sum_{k=1}^d \sigma(\bar{e}_{i,k}^{(u)} \cdot \bar{e}_{j,k}^{(v)})
\end{align}
where $s_{t,i,j}, \bar{s}_{i,j}\in\mathbb{R}$ denote the likelihood of $u_i$ interacting with $v_j$ in the $t$-th period and in long term, respectively. These predictions are made using the learned short-term collaborative embeddings $\textbf{e}_{t,i}^{(u)}, \textbf{e}_{t,j}^{(v)}$, and the learned long-term sequential embeddings $\bar{\textbf{e}}_i^{(u)}, \bar{\textbf{e}}_j^{(v)}$. And $e_{t,i,k}^{(u)}, e_{t,j,k}^{(v)}, \bar{e}_{i,k}^{(u)}, \bar{e}_{j,k}^{(v)}$ denote the element value of the $k$-th embedding dimension. The likelihood $s_{t,i',j'}$ and $\bar{s}_{i',j'}$ of user $u_{i'}$ interacting with item $v_{j'}$ are calculated in a similar manner. 

With the likelihood scores, \model\ aligns the score differences between the short and long-term views. Following the loss design of the main task, we adopt the SSL objective function:
\begin{align}
    \label{sal}
    &\mathcal{L}_{sal} = \sum_{t=1}^T\sum_{(u_i,v_j),(u_{i'},v_{j'})} \max(0, 1- d_1 \cdot d_2)\nonumber\\
    d_1 &= w_{t,i} \bar{s}_{i,j} - w_{t,i'} \bar{s}_{i',j'},~~~~
    d_2 = s_{t,i,j} - s_{t,i',j'}
\end{align}
where $d_1$ represents the likelihood difference between edge $(u_i, v_j)$ and $(u_{i'}, v_{j'})$ in the long-term view, and $d_2$ represents the difference score between edge $(u_i, v_j)$ and $(u_{i'}, v_{j'})$ in the $t$-th period short-term view. Specially, \model\ applies learned weights $w_{t,i}$ and $w_{t,i'}$ for capturing the different degrees of preference consistency between short-term and long-term for users. These weights, which are formally calculated as follows, personalize the self-supervised learning to avoid false cross-view preference alignment:
\begin{align}
    &w_{t,i} = \text{sigm} (\mathbf{\Gamma}_{t,i} \cdot \mathbf{W}_2+b_2),\nonumber\\
    \mathbf{\Gamma}_{t,i} = {\sigma}&( ( \bar{\e}^{(u)}_{i} + \e^{(u)}_{t,i} + \bar{\e}^{(u)}_{i} \odot \e^{(u)}_{t,i}) \mathbf{W}_1 + \mathbf{b}_1)
\end{align}
where $\mathbf{W}_1 \in \mathbb{R}^{d\times d_{sal}}$, $\mathbf{W}_2 \in \mathbb{R}^{d_{sal} \times 1}$, $\mathbf{b}_1 \in \mathbb{R}^{d_{sal}}$, $b_2 \in \mathbb{R}$ are trainable transformation parameters, sigm is sigmoid function. $w_{t, i}$ determines the stability of the long-term score $\bar{s}_{i,j}$ as a reference for noise correction in the $t$-th period.
We exclude the backpropagation of the long-term score $\bar{s}_{i,j}$ and $ \bar{s}_{i',j'}$, to direct the optimization process towards correcting the short-term score and learning the user stability weight $w_{t,i}$ and $w_{t,i'}$. In the loss function Eq. \eqref{sal}, the role of $d_1$ is to guide the optimization of $d_2$ in terms of direction and magnitude. Specifically, if $(u_i, v_j)$ exhibits a strong long-term relationship, while $(u_{i'}, v_{j'})$ has a weaker relationship (\ie, $\bar{s}_{i,j} > \bar{s}_{i',j'}$), and the preference stability of $u_i$ is higher than that of $u_{i'}$ (\ie, $w_{t,i} > w_{t,i'}$), then $d_1 > 0$. This leads to the optimization of $d_2$ towards a larger value, indicating that the short-term relationship $s_{t,i,j}$ corresponding to $(u_i, v_j)$ will be strengthened, while the short-term relationship $s_{t,i',j'}$ of $(u_{i'}, v_{j'})$ will be weakened. The larger the value of $d_1$, the greater the degree to which $d_2$ is optimized. A similar situation occurs when $d_1 < 0$, in which $d_2$ will be optimized for smaller negative values.

By combining the SSL task with the main recommendation task, we have the final loss function as follows:
\begin{equation}
    \mathcal{L}=\mathcal{L}_{rec}+\lambda_1\cdot \mathcal{L}_{sal} + \lambda_2 \cdot {\left\| \Theta \right\|}^2_{\rm{F}}
\end{equation}
where weights $\lambda_1$ and $\lambda_2$ are used to balance the importance of different loss terms. Additionally, we apply weight decay with $l_2$ regularization on the parameters $\Theta$ of our model. 

We summarize the learning process of \emph{\model} in Alg~\ref{alg:learn_alg}.

\begin{algorithm}[t]
\footnotesize
	\caption{Model Inference of the \model\ Framework.}
	\label{alg:learn_alg}
	\LinesNumbered
	\KwIn{number of short-term time intervals $T$, partitioned graph-structured interactions $\{\mathcal{A}_t|1\leq t\leq T\}$, user behavior sequences $\{v_{i,1},...,v_{i,m},...,v_{i,M}\}$, sample number $N$, maximum epoch number $E$, weight for loss of self-augmented learning $\lambda_1$, regularization weight $\lambda_2$, learning rate $\rho$} 
	\KwOut{trained parameters in $\mathbf{\Theta}$}
	Initialize all parameters in $\mathbf{\Theta}$\\
    \For{$e=1$ to $E$}{
        \For{$t=1$ to $T$}{
            Propagate multi-hop collaborative information through the LightGCN paradigm\\
            Obtain user short-term embeddings $\mathbf{e}_{t,i}^{(u)}$ and item short-term embeddings $\mathbf{e}_{t,j}^{(v)}$ at period $t$
        }
        Construct interval-level user feature change sequences $S_{i}^{intv}$ and item feature change sequences $S_{j}^{intv}$\\ 
        Generate interval-level long-term user embedding $\bar{\textbf{e}}_i^{(u)}$ and item embedding $\bar{\textbf{e}}_j^{(v)}$\\ 
        Capture instance-level sequential long-term user feature $\tilde{\textbf{e}}_i^{(u)}$\\
        Draw a mini-batch $\textbf{U}$ from all users, each with $N_{pr}$ positive-negative samples\\
        $\mathcal{L} = \lambda_2\cdot\|\mathbf{\Theta}\|_\text{F}^2$\\
        \For{each $u_i\in\textbf{U}$}{
            Compute predictions $\hat{\mathcal{A}}_{T+1,i,p_k}, \hat{\mathcal{A}}_{T+1, i, n_k}$\\
            $\mathcal{L}+=\sum_{k=1}^{N_{pr}}{\rm{max}}(0,1-\hat{\mathcal{A}}_{T+1,i,p_k}+\hat{\mathcal{A}}_{T+1, i, n_k})$\\
        }
        
        \For{$t=1$ to $T$}{
            Sample $N_{sal}$ pairs of user-item edge pairs $(u_i, v_j)$ and $(u_{i'}, v_{j'})$, denoted as $\textbf{P}$ \\
            \For{each $(u_i, v_j),(u_{i'}, v_{j'})\in\textbf{P}$}{
                Compute the likelihood score of user interacting with item in the $t$-th period, \ie, $s_{t,i,j}$ for $(u_{i}, v_{j})$ and $s_{t,i',j'}$ for $(u_{i'}, v_{j'})$ \\ 
                Compute the likelihood score of user interacting with item in long term, \ie,  $\bar{s}_{i,j}$ for $(u_{i}, v_{j})$ and $\bar{s}_{i',j'}$ for $(u_{i'}, v_{j'})$\\ 
                Compute user personalized weight $w_{t, i}$\\
                Compute long-term difference $d_1$ and short-term difference $d_2$ \\
                $\mathcal{L}+=\lambda_1 \cdot \max(0, 1- d_1 \cdot d_2)$
            }
        }
        \For{each parameter $\theta\in\mathbf{\Theta}$}{
            $\theta=\theta-\rho \cdot \partial \mathcal{L} / \partial \theta$
        }
    }
    return all parameters $\Theta$
\end{algorithm}
\subsection{In-Depth Analysis of \model }
\subsubsection{\bf Theoretical Analysis}
Our personalized SSL method addresses short-term noises by generating adaptive gradients based on users who exhibit significant disparities between their long-term and short-term preferences. More specifically, for Eq. \eqref{sal}, we can quantify the impact of edges $(u_i, v_j), (u_{i'}, v_{j'})$ in the short-term and long-term view on the node embeddings by:
\begin{align}
    &c'(w_{t,i'}) = \bar{s}_{i',j'} \cdot (s_{t,i,j} -s_{t,i',j'}) = \bar{s}_{i',j'} \cdot s_{t,i,j}-\bar{s}_{i',j'} \cdot s_{t,i',j'}\notag\\
    &c'(w_{t,i})=\bar{s}_{i,j} \cdot (s_{t,i',j'}-s_{t,i,j}) = \bar{s}_{i,j} \cdot s_{t,i',j'}-\bar{s}_{i,j} \cdot s_{t,i,j}
    \label{w}
\end{align}
\begin{align}
    c'(s_{t,i',j'})=w_{t,i} \cdot \bar{s}_{i,j}-w_{t,i'} \cdot \bar{s}_{i',j'}\notag\\
    c'(s_{t,i,j})=w_{t,i'} \cdot \bar{s}_{i',j'}-w_{t,i} \cdot \bar{s}_{i,j}
    \label{S}
\end{align}
where $c'(x)$ represents the gradient with respect to $x$. In Eq. \eqref{w}, if $s_{t,i,j}-s_{t,i',j'}>0$, the gradient $c'(w_{t,i'})$ will monotonically increase with the variation of $\bar{s}_{i',j'}$. Moreover, when the difference between the long-term value and short-term edge value is larger, \ie, when $\bar{s}_{i',j'} s_{t, i',j'}$ is smaller, the learning gradient of $w_{t, i'}$ will be larger. The same applies to the effect of $\bar{s}_{i,j}$ and $s_{t,i,j}$ on the gradient of $w_{t,i}$. In Eq. \eqref{S}, if $w_{t, i'}\bar{s}_{i',j'}$ is small, indicating that user $i'$ exhibits unstable short-term behavior and has a weak long-term relationship with item $j'$, the gradient of $s_{t, i',j'}$ will increase. Overall, self-augmented learning emphasizes gradients from samples with instability between long-term and short-term user-item relationships, thereby enhancing the model training process.
Additionally, short-term graphs offer a novel level of detail for understanding user behavior, thereby enhancing model performance.

\subsubsection{\bf Model Complexity Analyses}

The short-term GNN encoding takes $\sum_{t=1}^{T}O(L\times |\mathcal{A}_t|\times d)=O(L \times |\mathcal{A}|\times d)$ time complexity, where $|\mathcal{A}|$ is the number of all interaction edges, which means the time cost is the same as that of a complete long-term graph. The instance-level sequence learning takes $O((T \times d^2 + T^2 \times  d) \times (I+J))$, and interval-level sequence modeling with attention mechanism takes $O((M \times d^2 + M^2 \times  d) \times B)$, where $B$ is the batch size. In the SAL paradigm, the cost is $O(B \times N_{sal} \times d)$. we only need to do a vector dot product of randomly selected edges, requiring less time than commonly used SSL methods such as contrastive learning \cite{Oord_2018}.

%% file: eval.tex
\section{Evaluation}
\label{sec:eval}
Our experiments are designed to answer the following questions.

\noindent \textbf{RQ1}: How does \model\ perform \wrt\ top-\textit{k} recommendation as compared with the state-of-the-art models?

\noindent \textbf{RQ2}: What are the benefits of the components proposed?

\noindent \textbf{RQ3}: How does \model\ perform in the data noise issues?

\noindent \textbf{RQ4}: How do the key hyperparameters influence \model?

\noindent \textbf{RQ5}: In real cases, how can the designed self-augmented learning in \model\ provide useful interpretations?

\subsection{Experimental Settings}
\subsubsection{\bf Experimental Datasets} 
The experiments are conducted on four open-source datasets shown in Table \ref{tab:datasets}. \textbf{Amazon-book}~\cite{he2016ups}: it records users' ratings of Amazon books in 2014. \textbf{Gowalla}~\cite{cho2011friendship}: it is the user geolocation check-in dataset from Gowalla in 2010. \textbf{Movielens}~\cite{harper2015movielens}: it contains users' ratings for movies from 2002 to 2009. \textbf{Yelp}: this is a dataset of venue reviews, and we sample data from 2009 to 2019. We apply the 5-core setting to ensures that each user and item has at least five interactions.
\begin{table}[t]
    \small
    \centering
    \caption{Statistics of the experimental datasets.}
    \vspace{-10pt}
    \begin{tabular}{@{}lcccc@{}}
    \toprule
    Dataset     & User \# & Item \# & interaction \# & Density  \\ \midrule
    Amazon-book & 11199   & 30821   & 375916         & 1.1e-3 \\
    Gowalla     & 48653   & 52621   & 1807125        & 7.1e-4 \\
    Movielens   & 24312   & 8688    & 1758929        & 8.3e-3 \\ 
    Yelp        & 19751   & 38391   & 1467157        & 1.9e-3 \\
    \bottomrule
    \end{tabular}
    \label{tab:datasets}
    \vspace{-10pt}
\end{table}

\begin{table*}[t]
\centering
\small
\setlength{\tabcolsep}{3pt}
\caption{Performance comparison on Amazon, Gowalla, MovieLens and Yelp datasets in terms of \textit{HR} and \textit{NDCG}.}
\vspace{-10pt}
\begin{tabular}{|c|cccc|cccc|cccc|cccc|}
	\hline
	Data                    & \multicolumn{4}{c|}{Amazon}                                                            & \multicolumn{4}{c|}{Gowalla}                                                           & \multicolumn{4}{c|}{Movielens}                                                         & \multicolumn{4}{c|}{Yelp}                                                               \\ \hline
	\multirow{2}{*}{Metric} & \multicolumn{2}{c|}{Top 10}                          & \multicolumn{2}{c|}{Top 20}     & \multicolumn{2}{c|}{Top 10}                          & \multicolumn{2}{c|}{Top 20}     & \multicolumn{2}{c|}{Top 10}                          & \multicolumn{2}{c|}{Top 20}     & \multicolumn{2}{c|}{Top 10}                          & \multicolumn{2}{c|}{Top 20}      \\ \cline{2-17} 
							& HR             & \multicolumn{1}{c|}{NDCG}             & HR             & NDCG             & HR             & \multicolumn{1}{c|}{NDCG}             & HR             & NDCG             & HR             & \multicolumn{1}{c|}{NDCG}             & HR             & NDCG             & HR             & \multicolumn{1}{c|}{NDCG}             & HR             & NDCG             \\ \hline
	BiasMF                  & 0.314          & \multicolumn{1}{c|}{0.170}          & 0.457          & 0.206          & 0.543          & \multicolumn{1}{c|}{0.362}          & 0.669          & 0.393          & 0.208          & \multicolumn{1}{c|}{0.115}          & 0.311          & 0.141          & 0.281          & \multicolumn{1}{c|}{0.151}          & 0.411          & 0.184          \\
	NCF                     & 0.306          & \multicolumn{1}{c|}{0.165}          & 0.431          & 0.197          & \underline{0.606}          & \multicolumn{1}{c|}{\underline{0.418}}          & \underline{0.716}          & \underline{0.446}          & 0.198          & \multicolumn{1}{c|}{0.107}          & 0.299          & 0.133          & 0.313          & \multicolumn{1}{c|}{0.169}          & 0.462          & 0.206          \\
	GRU4Rec                 & 0.148          & \multicolumn{1}{c|}{0.077}          & 0.232          & 0.098          & 0.393          & \multicolumn{1}{c|}{0.253}          & 0.515          & 0.284          & 0.127          & \multicolumn{1}{c|}{0.076}          & 0.238          & 0.099          & 0.087          & \multicolumn{1}{c|}{0.043}          & 0.143          & 0.057          \\
	SASRec                  & 0.268          & \multicolumn{1}{c|}{0.171}          & 0.324          & 0.185          & 0.562          & \multicolumn{1}{c|}{0.360}          & 0.688          & 0.392          & 0.121          & \multicolumn{1}{c|}{0.072}          & 0.163          & 0.082          & 0.106          & \multicolumn{1}{c|}{0.057}          & 0.151          & 0.057          \\
	TiSASRec                & 0.270          & \multicolumn{1}{c|}{0.171}          & 0.326          & 0.185          & 0.573          & \multicolumn{1}{c|}{0.376}          & 0.690          & 0.405          & 0.120          & \multicolumn{1}{c|}{0.071}          & 0.165          & 0.082          & 0.092          & \multicolumn{1}{c|}{0.051}          & 0.133          & 0.061          \\
	Bert4Rec                & 0.312          & \multicolumn{1}{c|}{0.173}          & 0.445          & 0.207          & 0.544          & \multicolumn{1}{c|}{0.359}          & 0.676          & 0.393          & 0.175          & \multicolumn{1}{c|}{0.087}          & 0.299          & 0.118          & 0.290          & \multicolumn{1}{c|}{0.158}          & 0.428          & 0.193          \\
	NGCF                    & 0.277          & \multicolumn{1}{c|}{0.147}          & 0.307          & 0.180          & 0.550          & \multicolumn{1}{c|}{0.347}          & 0.692          & 0.383          & 0.147          & \multicolumn{1}{c|}{0.076}          & 0.238          & 0.099          & 0.325          & \multicolumn{1}{c|}{0.174}          & 0.475          & 0.212          \\
	LightGCN                & 0.244          & \multicolumn{1}{c|}{0.127}          & 0.271          & 0.159          & 0.456          & \multicolumn{1}{c|}{0.286}          & 0.592          & 0.318          & 0.211          & \multicolumn{1}{c|}{0.115}          & 0.319          & 0.142          & \underline{0.346}          & \multicolumn{1}{c|}{\underline{0.189}}          & \underline{0.495}          & \underline{0.226}          \\
	SRGNN                   & 0.199          & \multicolumn{1}{c|}{0.109}          & 0.281          & 0.129          & 0.575          & \multicolumn{1}{c|}{0.371}          & 0.687          & 0.399          & 0.122          & \multicolumn{1}{c|}{0.069}          & 0.166          & 0.080          & 0.097          & \multicolumn{1}{c|}{0.054}          & 0.137          & 0.064          \\
	GCE-GNN                 & 0.279          & \multicolumn{1}{c|}{0.145}          & 0.401          & 0.180          & 0.578          & \multicolumn{1}{c|}{0.373}          & 0.709          & 0.406          & 0.169          & \multicolumn{1}{c|}{0.088}          & 0.264          & 0.112          & 0.297          & \multicolumn{1}{c|}{0.157}          & 0.440          & 0.194          \\
	SURGE                   & \underline{0.362}          & \multicolumn{1}{c|}{0.215}          & \underline{0.468}          & \underline{0.242}          & 0.479          & \multicolumn{1}{c|}{0.259}          & 0.654          & 0.304          & \underline{0.237}          & \multicolumn{1}{c|}{0.124}          & \underline{0.337}          & \underline{0.149}          & 0.278          & \multicolumn{1}{c|}{0.144}          & 0.428          & 0.182          \\
	SGL                     & 0.336          & \multicolumn{1}{c|}{0.184}          & 0.465          & 0.217          & 0.413          & \multicolumn{1}{c|}{0.259}          & 0.495          & 0.280          & 0.223          & \multicolumn{1}{c|}{0.120}          & 0.282          & 0.135          & 0.320          & \multicolumn{1}{c|}{0.170}          & 0.461          & 0.206          \\
	DGCF                     & 0.307          & \multicolumn{1}{c|}{0.163}          & 0.357          & 0.201         & 0.524          & \multicolumn{1}{c|}{0.339}          & 0.629          & 0.388          & 0.216          & \multicolumn{1}{c|}{0.116}          & 0.309          & 0.136          & 0.318          & \multicolumn{1}{c|}{0.168}          & 0.452          & 0.202          \\
	TGSRec                     & 0.228          & \multicolumn{1}{c|}{0.172}          & 0.426          & 0.289          & 0.424          & \multicolumn{1}{c|}{0.347}          & 0.695          & 0.557          & 0.145          & \multicolumn{1}{c|}{0.125}          & 0.229          & 0.177          & 0.217          & \multicolumn{1}{c|}{0.143}          & 0.458          & 0.301          \\
	ICLRec                  & 0.285          & \multicolumn{1}{c|}{\underline{0.226}}          & 0.343          & 0.240          & 0.197          & \multicolumn{1}{c|}{0.122}          & 0.265          & 0.193          & 0.190          & \multicolumn{1}{c|}{\underline{0.126}}          & 0.248          & 0.140          & 0.195          & \multicolumn{1}{c|}{0.113}          & 0.274          & 0.133          \\
	CoSeRec                 & 0.368          & \multicolumn{1}{c|}{0.230}          & 0.465          & 0.254          & 0.556          & \multicolumn{1}{c|}{0.362}          & 0.669          & 0.390          & 0.166          & \multicolumn{1}{c|}{0.112}          & 0.221          & 0.125          & 0.195          & \multicolumn{1}{c|}{0.110}          & 0.270          & 0.129          \\
	CoTRec                  & 0.338          & \multicolumn{1}{c|}{0.213}          & 0.421          & 0.233          & 0.531          & \multicolumn{1}{c|}{0.400}          & 0.608          & 0.419          & 0.216          & \multicolumn{1}{c|}{0.117}          & 0.311          & 0.141          & 0.283          & \multicolumn{1}{c|}{0.152}          & 0.416          & 0.185          \\
	CLSR                    & 0.332          & \multicolumn{1}{c|}{0.189}          & 0.441          & 0.217          & 0.529          & \multicolumn{1}{c|}{0.296}          & 0.699          & 0.339          & 0.061          & \multicolumn{1}{c|}{0.036}          & 0.076          & 0.040          & 0.261          & \multicolumn{1}{c|}{0.134}          & 0.400          & 0.169          \\
    \hline
	\textit{\model}                    & \textbf{0.391} & \multicolumn{1}{c|}{\textbf{0.240}} & \textbf{0.487} & \textbf{0.264} & \textbf{0.637} & \multicolumn{1}{c|}{\textbf{0.437}} & \textbf{0.745} & \textbf{0.465} & \textbf{0.239} & \multicolumn{1}{c|}{\textbf{0.128}} & \textbf{0.341} & \textbf{0.154} & \textbf{0.365} & \multicolumn{1}{c|}{\textbf{0.201}} & \textbf{0.509} & \textbf{0.237} \\ \hline
	p-val.                  & $2.9e^{-4}$    & \multicolumn{1}{c|}{$2.7e^{-3}$}    & $8.4e^{-5}$    & $3.1e^{-3}$    & $5.6e^{-6}$    & \multicolumn{1}{c|}{$9.5e^{-5}$}    & $2.0e^{-8}$    & $2.0e^{-8}$    & $3.1e^{-4}$    & \multicolumn{1}{c|}{$1.7e^{-3}$}    & $8.7e^{-4}$    & $5.5e^{-4}$    & $1.8e^{-4}$    & \multicolumn{1}{c|}{$2.2e^{-4}$}    & $2.3e^{-5}$    & $7.6e^{-5}$    \\ \hline
\end{tabular}

\label{tab:overall_performance}
\end{table*}
\subsubsection{\bf Evaluation Protocols} We follow the evaluation protocol of recent works on sequential recommendation~\cite{ICLRec,CoSeRec}. The data is split into three sets: the most recent interaction of each user is used as the test set, the penultimate one is used as the validation set, and the remaining interactions in users' sequences are used as the training data. We randomly sample 10,000 users as test users. For each test user, we sample negative samples by randomly selecting 999 items that the user does not interact with. For evaluation, both positive and negative items are ranked together for each user and we employ two metrics, \ie, \textit{Hit Rate (HR)@N} and \textit{Normalized Discounted Cumulative Gain (NDCG)@N}~\cite{Ren_2020,NGCF}.

\subsubsection{\bf Compared Baseline Methods} We compare our \model\ with state-of-the-art baselines from different research lines. 

\noindent \textbf{Conventional Factorization-based Technique.}
\begin{itemize}[leftmargin=*]
\item \textbf{BiasMF} \cite{BMF}: It augments matrix factorization to incorporate implicit feedback, temporal effects, and confidence levels.
\end{itemize}

\noindent \textbf{Neural Factorization Method.}
\begin{itemize}[leftmargin=*]
\item \textbf{NCF} \cite{NCF}: This method designs a neural network framework for collaborative filtering based on neural networks.
\end{itemize}

\noindent \textbf{Sequential Recommendation.}
\begin{itemize}[leftmargin=*]
\item \textbf{GRU4Rec} \cite{GRU4Rec}: It is a session-based recommendation with RNN.
\item \textbf{SASRec} \cite{SASRec}: The model employs self-attention mechanisms to capture the sequential patterns within recommender systems.
\item \textbf{TiSASRec} \cite{TiSASRec}: 
This method proposes to view the user's interactions as a sequence with different time intervals, modeling them as relations between any two interactions.
\item \textbf{Bert4Rec} \cite{Bert4Rec}: This approach models user behavior sequences with the bidirectional self-attention network based on Transformer architecture through the Cloze task.
\end{itemize}

\noindent \textbf{Traditional Graph Neural Networks Method.}
\begin{itemize}[leftmargin=*]
\item \textbf{NGCF} \cite{NGCF}: This method crafts a graph neural network to enhance high-order collaborative filtering in recommendation.
\item \textbf{LightGCN} \cite{LightGCN}: It simplifies the GCNs in recommendation.
\item \textbf{SRGNN} \cite{SRGNN}: It model separated session sequences into graph structure to capture complex item transitions.
\item \textbf{GCE-GNN} \cite{GCE-GNN}: It introduces the global-level and the session-level pairwise item-transition graph for sequence recommender.
\end{itemize}

\noindent \textbf{Dynamic or Temporal Graph Neural Networks Method.}
\begin{itemize}[leftmargin=*]
\item \textbf{SURGE} \cite{SURGE}: It proposes to aggregate implicit signals into explicit ones from user behaviour sequences by designing GNN models. 
\item \textbf{DGCF} \cite{li2020dynamic}: It utilizes dynamic graph structures to effectively encapsulate the collaborative and sequential dynamics.
\item \textbf{TGSREC} \cite{fan2021continuous}: It proposes a Temporal Collaborative Transformer layer that integrates collaborative attention to capture both user-item interactions and temporal dynamics over a temporal graph.
\end{itemize}

\noindent \textbf{Recommenders enhanced by Self-Supervised Learning.}
\begin{itemize}[leftmargin=*]
\item  \textbf{SGL} \cite{SGL}: This model employs random techniques for structural and feature augmentation, creating diverse data perspectives that enrich the self-supervised learning process.
\item \textbf{ICLRec} \cite{ICLRec}:  The model discerns user intents by clustering across all user behavior sequences and refines these intents through contrastive learning to ensure alignment with user expectations. 
\item \textbf{CoSeRec} \cite{CoSeRec}: It introduces new sequential data augmentation strategies to enhance contrastive learning for recommendation.
\item \textbf{CoTRec} \cite{CoTRec}: It proposes a self-supervised method to exploit the session graph to two views and two distinct graph encoders.
\item \textbf{CLSR} \cite{CLSR}: It is a contrastive learning method to capture long and short-term interests by comparing with proxy representations.
\end{itemize}
\subsubsection{\bf Implementation Details} 
We implement \model\ with TensorFlow and use Adam optimizer with the $1e^{-3}$ learning rate and 0.96 epoch decay ratio. The embedding dimension size is 64. The number of graph neural layers is selected from \{1,2,3\}, the training batch size is selected from \{128, 256, 512\}, the attention layer for instance-level sequence is selected from \{2,3,4\}, the short-term graph number $T$ is selected from $\{t|2\leq t \leq 14\}$, the dimension of personalized weight is selected from \{16,32,48\}, the weight $\lambda_1$ for SSL loss is tuned from $\{1e^{-4},1e^{-5},1e^{-6},1e^{-7}\}$, the weight $\lambda_2$ for regularization loss is $1e^{-2}$, and dropout rate is 0.5. 
\subsection{Overall Performance Comparison (RQ1)}
The results are presented in Table \ref{tab:overall_performance}. \model\ and the best-performed baseline are retrained 5 times for p-values.
\begin{itemize}[leftmargin=*]
\item  \textbf{Overall Performance Validation.}
Our proposed \model\ consistently outperforms all other SOTAs across various evaluation metrics. This observation substantiates the superiority of \model, attributed to the following factors. i) Integration of short-term graphs and long-term sequence information: Our model leverages both short-term collaborative information and long-term holistic dependencies to capture user interest changes. By combining these two types of information, our model gains a comprehensive understanding of user preferences and achieves improved performance. ii) Personalized self-augmented learning schema: Our model benefits from a personalized self-augmented learning framework, which enables it to handle the noise present in sparse short-term user interactions. Through cross-view supervision signals that bridge the gap between short-term and long-term representations, our model can effectively correct for noise and enhance the quality of recommendations. In addition, note that the Gowalla dataset is a locally-punched dataset, with many users repeatedly going to certain locations and checking in, so the simple NCF model performed best in these baselines.\\\vspace{-0.12in}

\item \textbf{Superiority of Sequential Recommendation with Graph.} Referring to state-of-the-art baselines, sequential approaches based on dynamic graphs such as DGCF, TGSRec, and SURGE outperform most conventional Attention-based and RNN-based models (\eg, SASRec, TiSASRec, Bert4Rec, and GRU4Rec), as graphs can extract collaborative signals instead of solely modeling the sequence-specific context. Meanwhile, our \model\ transforms long-term information into multiple short-period graphs, which aggregate time and collaborative signals at different stages, reflecting the challenging long-range dependence of users through stage preferences. Consequently, \model\ exhibits superior performance compared to traditional GNN-based recommendations and sequential recommendations.\\\vspace{-0.12in}

\item \textbf{Effectiveness of Self-Supervised Learning.} From the evaluation results, it is evident that self-supervised learning significantly enhances the performance of all approaches (\eg, SGL, ICLRec, CoSeRec, CoTRec, and CLSR), regardless of whether they are GNN-based or sequence-based. TGSRec and our \model\ similarly combine collaborative filtering signals with attention-based sequential learning. However, \model\ significantly outperforms TGSRec, thanks to our self-supervised learning paradigm, which TGSRec does not incorporate for model enhancement. Self-supervised learning leverages different views of the data itself to augment supervision signals, thereby addressing the crucial limitation of supervision insufficiency in recommendation. Specifically, SGL employs stochastic graph data augmentation. ICLRec and CoSeRec generate self-augmented sequences through insertion and substitution operations. CoTRec formulates contrastive learning as maximizing agreement between the representation of the last clicked item and the predicted item. CLSR adopts a contrastive task to supervise the similarity between long and short-term interest and their corresponding interest proxies. In comparison to the aforementioned methods, the self-augmentation in our \model\ offers two main advantages: i) Instead of relying on random masking or permutation, which may inadvertently introduce noises into the sequence and collaboration modeling, our approach is based on users' short-term and long-term preferences using the original data. ii) We propose an innovative denoising SSL approach that explicitly uses stable long-term patterns to filter users' random and noisy short-term behaviors. This allows for the removal of noise caused by abnormal short-term interactions of stable users. 
\end{itemize}
\subsection{Ablation Study of \model\ (RQ2)}
\begin{table}[t]
    \centering
    \small
    \setlength{\tabcolsep}{1pt}
        \caption{Module ablation study on \model~\wrt~top 10.}
        \vspace{-10pt}
\begin{tabular}{cc|cc|cc|cc|cc}
\hline
\multicolumn{1}{c|}{\multirow{2}{*}{Category}}  & Data          & \multicolumn{2}{c|}{Amazon}     & \multicolumn{2}{c|}{Gowalla}    & \multicolumn{2}{c|}{Movielens}  & \multicolumn{2}{c}{Yelp}        \\ \cline{2-10} 
\multicolumn{1}{c|}{}                           & Variants      & HR             & NDCG           & HR             & NDCG           & HR             & NDCG           & HR             & NDCG           \\ \hline
\multicolumn{1}{c|}{\multirow{2}{*}{SAL}}       & \textit{-SAL} & 0.364          & 0.218          & 0.631          & 0.428          & 0.211          & 0.110          & 0.328          & 0.173          \\
\multicolumn{1}{c|}{}                           & \textit{-UW}  & 0.355          & 0.219          & 0.635          & 0.430          & 0.198          & 0.104          & 0.268          & 0.136          \\ \hline
\multicolumn{1}{c|}{Short-term}                 & \textit{-STG} & 0.347          & 0.206          & 0.525          & 0.353          & 0.165          & 0.088          & 0.313          & 0.171          \\ \hline
\multicolumn{1}{c|}{\multirow{3}{*}{Long-term}} & \textit{-ATL} & 0.337          & 0.200          & 0.564          & 0.363          & 0.211          & 0.115          & 0.324          & 0.175          \\
\multicolumn{1}{c|}{}                           & \textit{-GAT} & 0.277          & 0.164          & 0.541          & 0.339          & 0.185          & 0.097          & 0.248          & 0.134          \\
\multicolumn{1}{c|}{}                           & \textit{-CF}  & 0.247          & 0.156          & 0.320          & 0.209          & 0.170          & 0.089          & 0.246          & 0.133          \\ \hline
\multicolumn{2}{c|}{\emph{\model}}                              & \textbf{0.387} & \textbf{0.239} & \textbf{0.637} & \textbf{0.437} & \textbf{0.239} & \textbf{0.128} & \textbf{0.365} & \textbf{0.201} \\ \hline
\end{tabular}
    \label{tab:module_ablation}
    \vspace{-10pt}
\end{table}
We individually remove the applied techniques from three major parts for ablation study, and Table \ref{tab:module_ablation} shows the results.
\begin{itemize}[leftmargin=*]
\item  \textbf{Effect of Short-term Graph Structure Learning.} We replace the several short-term graphs with a global user-item graph, denoted as \textit{-STG}. From the table, it is evident that the \textit{-STG} model significantly reduces the accuracy of the recommendation task. This is attributed to that our \model\ captures CF relationships and their changes at different periods and emphasizes the more important stages through temporal attention. In contrast, the static graph used in \textit{-STG} fails to capture these periodic aspects.\\\vspace{-0.12in}
\item  \textbf{Effect of Long-term Learning.} In the long-term feature representation, we construct a variant called \textit{-GAT} by replacing the GRU-attention mechanism for interval-level learning with a feature fusion method based on summation and another variant called \textit{-ATL} by removing the attention-based instance-level sequential pattern. It is evident from the results that \model\ consistently outperforms the variants. Assigning different importance to user preferences at different stages is crucial, as the GRU effectively captures the position information and dynamic interest changes. Additionally, attention modeling for long-term behavior at the instance level represents users from a comprehensive sequential perspective, allowing for fusion and mutual supervision with interval-level features from short-term graphs. Regarding the effectiveness of collaborative relation encoding, we conducted ablation experiments that remove the collaborative filtering paradigm, as indicated by \textit{-CF}. It demonstrate that constructing the user-item collaborative graph is crucial for the sequence recommendation. The collaborative knowledge it acquires enhances the model's ability to represent users.\\\vspace{-0.12in}
\item  \textbf{Effect of Long and Short-term Personalized Self-Augmented Learning.} We also conduct ablations on the components of personalized self-augmented learning by removing the personalized user weight (\textit{-UW}) and the entire self-augmented learning schema (\textit{-SAL}). The experimental results prove that supervising short-term graph noise through long-term features significantly improves the recommendation effectiveness. Comparing the results of \textit{-UW} and \model, it can be observed that weighting for different users is crucial for the Amazon, Yelp and Movielens datasets, in which many users may have variable behavioral interests so that personalized weights are needed for discrimination of noise and selective correction of short-term behavior encoding.
\end{itemize}

\subsection{Model Robustness Test (RQ3)}
\subsubsection{\bf Performance \wrt\ Data Noise Degree}
\begin{figure}[t]
    \centering
    \subfigure[Amazon data]{
        \includegraphics[width=0.48\columnwidth]{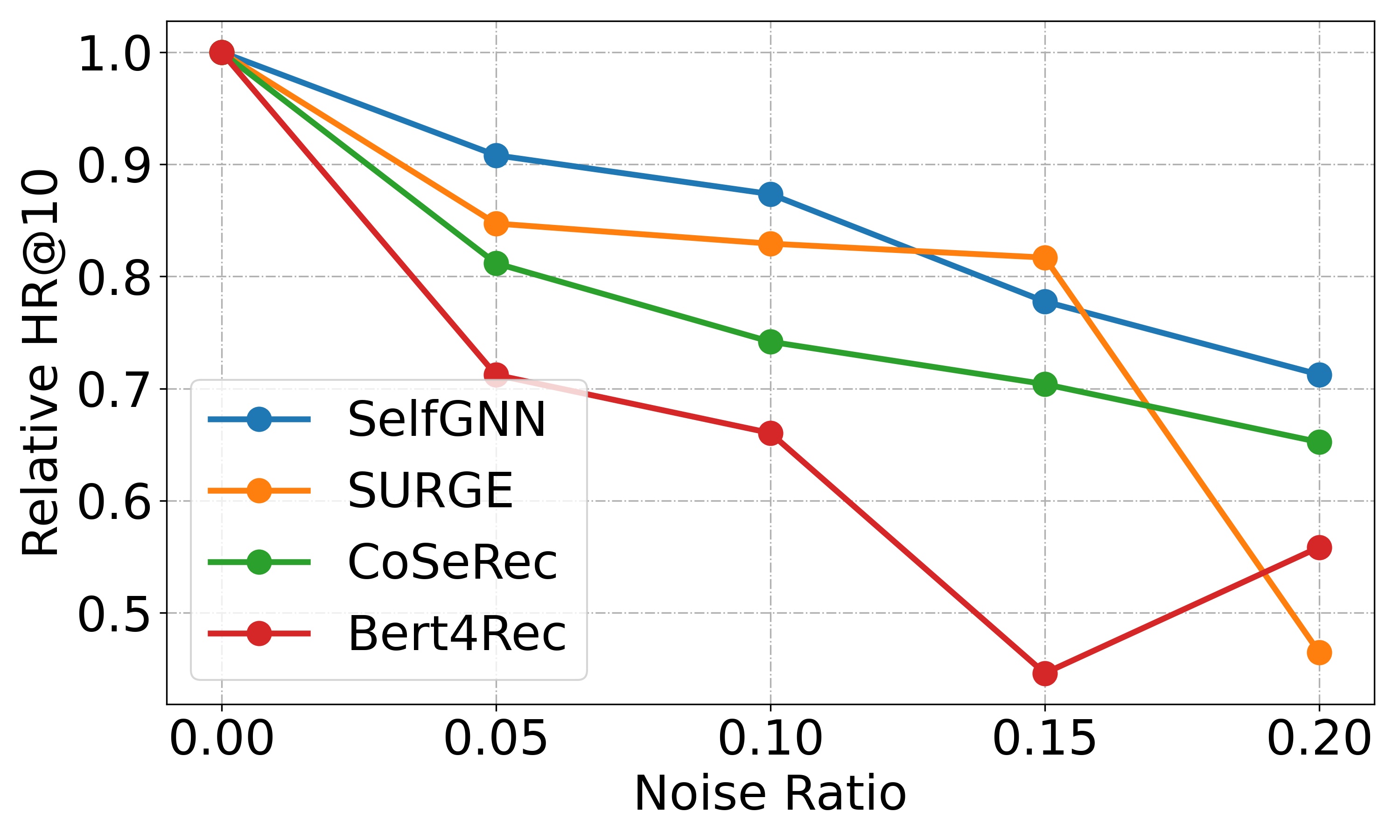}\ 
        \includegraphics[width=0.48\columnwidth]{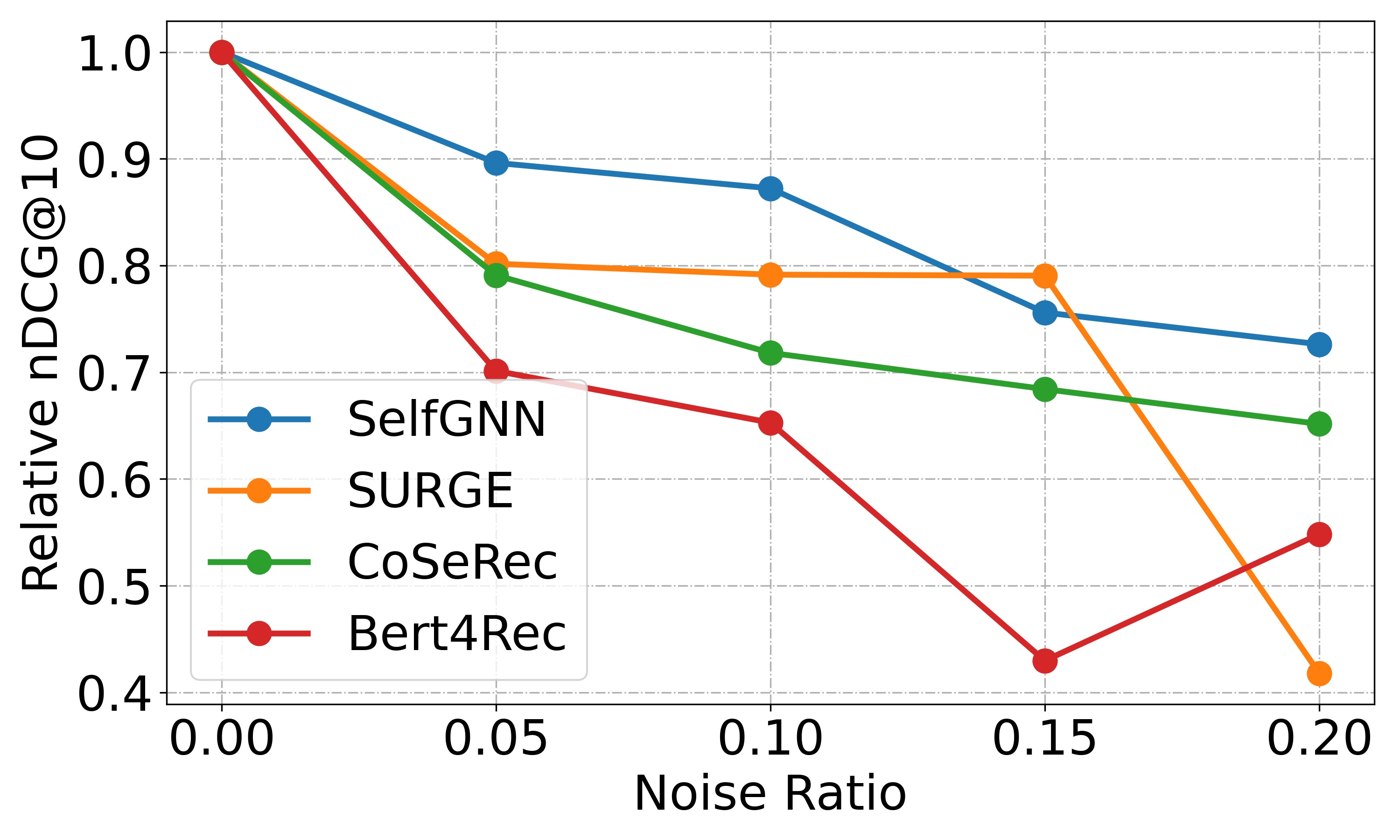}
        \vspace{-0.15in}
    }
    \subfigure[Movielens data]{
        \includegraphics[width=0.48\columnwidth]{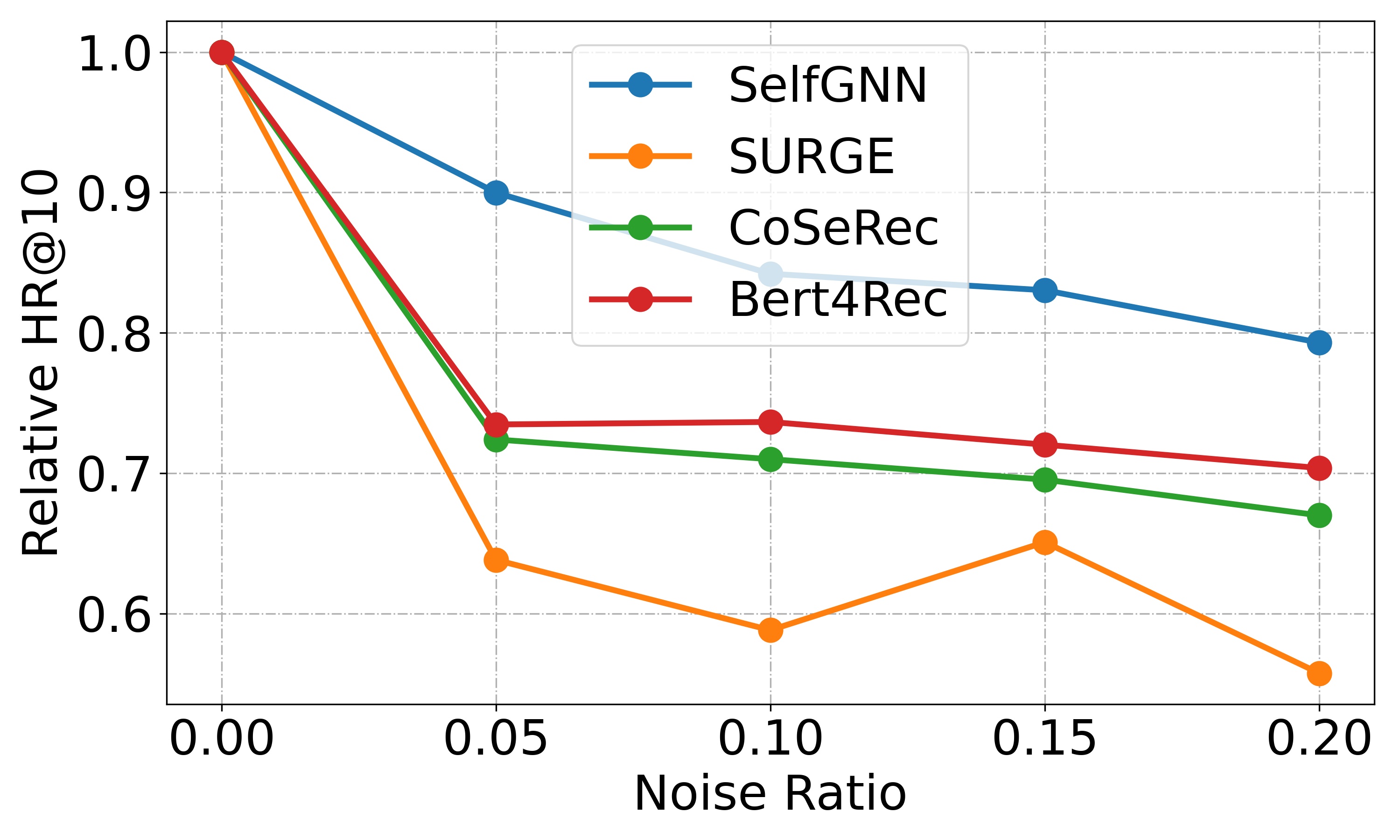}\ 
        \includegraphics[width=0.48\columnwidth]{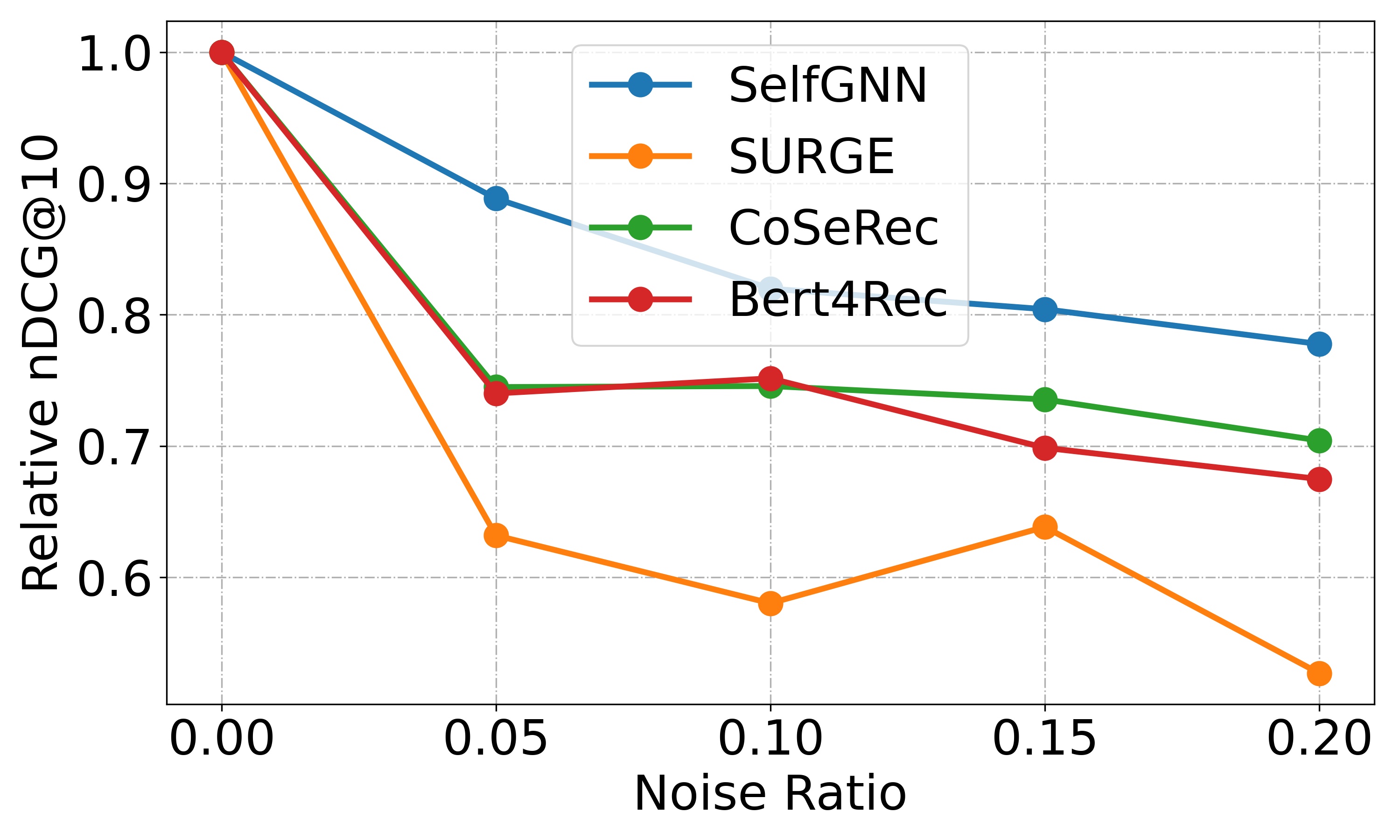}
    }
    \vspace{-0.15in}
    \caption{Relative performance degradation \wrt\ noise ratio.}
    \vspace{-0.2in}
    \label{fig:noise}
\end{figure}
To evaluate the robustness of \model\ against noise issues, we conduct experiments by randomly replacing different percentages of real interaction items with randomly-generated fake items for all users and retraining the model using the corrupted sequences as input. The noise ratios considered are 5\%, 10\%, 15\%, and 20\%, respectively. The target models for evaluation are SURGE, CoSeRec, and Bert4Rec, which are well-performing sequential recommendation models. Figure \ref{fig:noise} illustrates the performance degradation in different noise scenarios, showcasing the potential of our \model\ in addressing data noise. In the sparser Amazon dataset, while SURGE initially performs well with small amounts of noise, its performance rapidly declines when faced with 20\% noise. On the other hand, our \model\ is less affected by noise compared to the other models. Even with 20\% noise, it achieves a \textit{HR@10} value of 72\% and an \textit{NDCG@10} value of 73\% in the absence of noise. In the relatively denser Movielens dataset, our model attains a relative \textit{HR@10} value of 79\% and an \textit{NDCG@10} value of 78\% in the case of 20\% noise. We attribute this superiority to \model's ability to mitigate noise in the short-term graph through long-term features in personalized self-augmented learning. Furthermore, our long-term representation is obtained by combining short-term features through interval-level GRU attention, allowing the final long-term user interest representation to benefit from the denoising effect when short-term noise is reduced.
\subsubsection{\bf Performance \wrt\ Data Sparsity}
\begin{figure}
    \centering
    \subfigure[Amazon data]{
        \includegraphics[width=0.47\columnwidth]{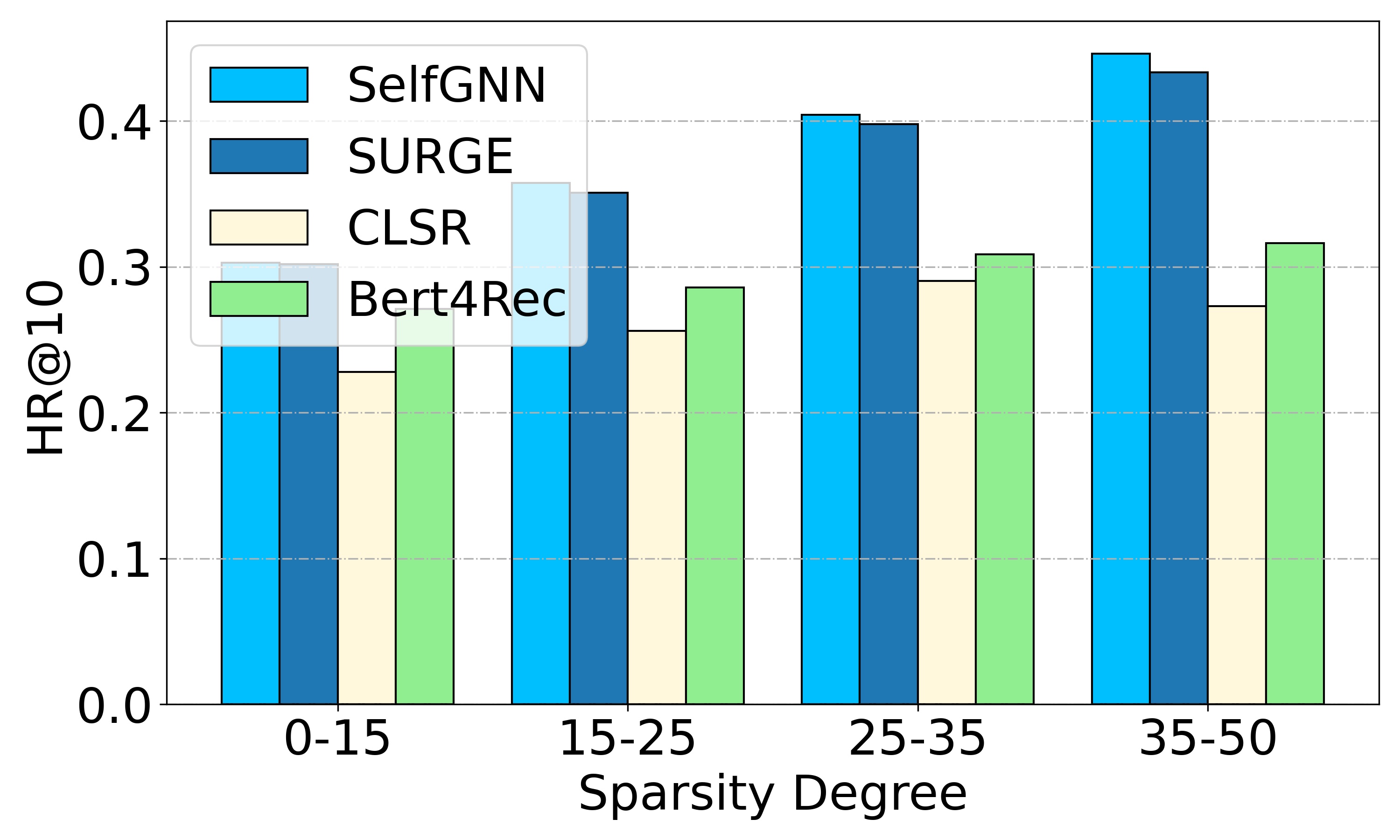}
    }
    \subfigure[Gowalla data]{
        \includegraphics[width=0.47\columnwidth]{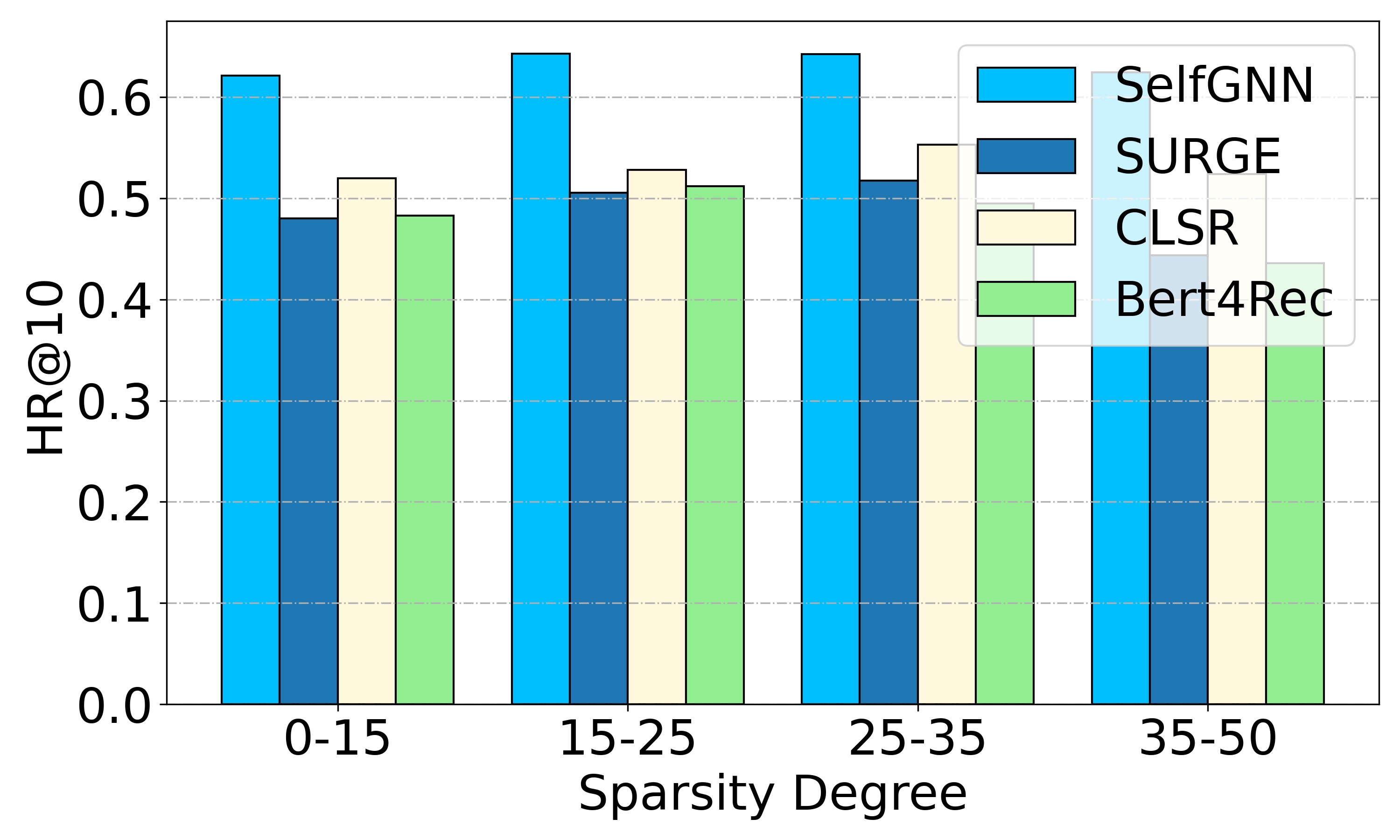}
    }
    \vspace{-0.15in}
    \caption{Performance \wrt\ interaction degrees.}
    \vspace{-0.1in}
    \label{fig:sparsity_amazon}
\end{figure}
To assess the impact of data sparsity on model performance, we categorize users into different groups based on their interaction numbers (\eg, 0-15, 15-25) and evaluate our \model\ alongside the top-performing models (\ie, SURGE, CLSR, and Bert4Rec). As depicted in Figure \ref{fig:sparsity_amazon}, the models generally exhibit lower predictive accuracy for sparser users. However, our \model\ consistently demonstrates greater stability than the other models. This can be attributed to our approach, which combines collaborative graph modeling and attention-based sequence modeling, compensating for missing information and leveraging additional self-supervised signals. In contrast, SURGE relies solely on item-item graphs for modeling, which may not capture sufficient information in sparse scenarios. This observation is further corroborated by the experimental results from the Gowalla dataset. Similarly, the sequential models (\ie, CLSR and Bert4Rec) also exhibit poorer performance when the sequences are very short.
\subsection{Hyperparameter Analysis (RQ4)}
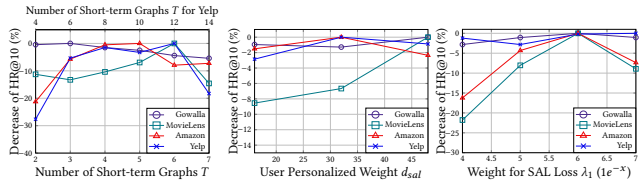
\begin{figure}
    \centering
    \begin{adjustbox}{max width=1.0\linewidth}
    \input{./figure/hyper1.tex}
    \end{adjustbox}
    \vspace{-0.2in}
    \caption{Hyperparameter study of the \model.}
    \vspace{-0.15in}
    \label{fig:hyperparam}
\end{figure}

The decrease of the experimental results with the changing of two key hyperparameters (\ie, $T$, $d_{sal}$ and $\lambda_1$) is plotted in Figure \ref{fig:hyperparam}. Firstly, when the dimension $d_{sal}$ for user personalization weight is set to 48 on the Gowalla and Movielens and is set to 32 on Amazon and Yelp, \model\ achieves optimal performance. Secondly, the performance initially improves and then declines as the number of short-term graph partitions $T$ increases because a higher number of partitions allows the model to capture more fine-grained information and if $T$ becomes too large, the graphs become excessively sparse. The results shows the optimal number of splits varies across different datasets due to their duration and densities. For instance, the optimal number of short-term partitions for the Yelp dataset is 12, while for Gowalla, only $T=3$ are needed. Thirdly, the best value for the self-augmented loss weight $\lambda_1$ is $1e^{-7}$ for the Yelp dataset and $1e^{-6}$ for the rest of the dataset. 
\subsection{Case Study (RQ5)}
\begin{figure}[t]
    \centering
    \includegraphics[width=0.42\textwidth]{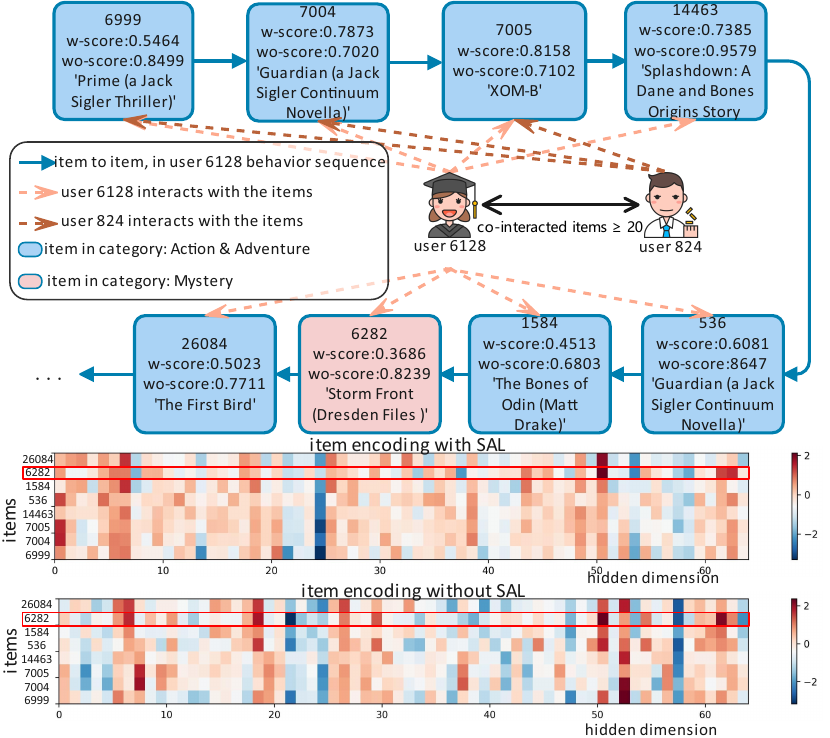}
    \vspace{-10pt}
    \caption{User-item likelihood scores in a partial sequence of user (6128) and heatmap of items' embedding in \model.}
    \vspace{-10pt}
    \label{fig:casestudy}
\end{figure}

In this section, we utilize concrete data examples to investigate the effect of self-augmented learning denoising. As shown in Figure \ref{fig:casestudy}, we randomly selected a user (6128) and part of the behavior sequence of that user, and the other user (824) which has at least 20 interaction items in common with the user (6128). We calculate and normalize the final user-item similarity scores $\hat{\mathcal{A}}_{T+1,i,j}$ in the case of both without (wo-score) and with (w-score) self-augmented learning. We display the item's score, title, and category in Figure \ref{fig:casestudy}. Upon examination, it can be observed that the score for item (6282) decreases from 0.8239 to 0.3686 with the inclusion of self-augmented learning. This indicates that the model identifies it as a noisy interaction that needs to be weakened. We prove that this behavior may be noise behavior from two aspects. Firstly, the category of item (6282) is "Mystery" which differs from the category of other items ("Action \& Adventure") that the user follows. Additionally, as a user who shares numerous similar interests with user (6128), user (824) does not follow item (6282). Moreover, from the encoding heatmap of items, it is evident that item (6282) exhibits distinctive features across multiple dimensions compared to other items. Furthermore, the features of items interacted with by the same user in a continuous manner exhibit greater dissimilarity compared to cases where SAL is not employed, which proves the SAL paradigm lightens the smoothing problem caused by GNNs.

To validate whether other noise items similar to item (6282) in all user sequences exhibit significant feature differences compared to the remaining normal items in their respective sequences, we conducted a statistical experiment. We computed the average cosine similarity of the feature embeddings between item (6282) and other items in the behavior sequence of user (6128) under both with and without self-augmented paradigms. Then, we similarly calculated the average cosine similarity between items that satisfy the noise conditions in other user sequences and the remaining items in the sequence where this item is located. The statistical results are presented in Table \ref{tab:case}, which show that in the self-augmented learning model (w-SAL), the average cosine similarity between noisy items and other items is significantly lower than the corresponding value in the model without self-augmented learning (w/o-SAL). In summary, our model demonstrates the ability to accurately identify and weaken interaction behaviors that are likely to be noise.
\begin{table}[h]
    \centering
    \small
    \setlength{\tabcolsep}{1.5pt}
\caption{Similarity statistics of noise items with other items.}    
\vspace{-10pt}
\begin{tabular}{c|cc|cc}
\hline
User                      & \multicolumn{2}{c|}{User 6218}      & \multicolumn{2}{c}{Other Users} \\ \hline
Model                     & \multicolumn{1}{c|}{w-SAL} & w/o-SAL & \multicolumn{1}{c|}{w-SAL} & w/o-SAL \\ \hline
Average Cosine Similarity & \multicolumn{1}{c|}{0.605} & 0.715  & \multicolumn{1}{c|}{0.620} & 0.711  \\ \hline
\end{tabular}
\vspace{-10pt}
\label{tab:case}
\end{table}

%% file: figure/hyper1.tex
\begin{tikzpicture}

\begin{axis}[
    width=12cm, 
    height=9cm, 
    xlabel={Number of Short-term Graphs $T$},
    ylabel={Decrease of HR@10 (\%)},
    xmin=2,xmax=7,
    ymin=-40,ymax=5,
    xtick distance=1, 
    label style={scale=2.5},
    legend columns=1,
    legend cell align=right,
    grid=both,
    every axis plot/.append style={ultra thick},
    every tick label/.append style={scale=1.7},
    legend style={
        nodes={scale=1.5, transform shape},
        legend image post style={scale=1.5},
        },
    legend style={at={(1,0)},anchor=south east},
    every axis plot post/.append style={
        every mark/.append style={scale=2}
    },
]
\addplot[color={rgb:red,133;green,76;blue,255}, mark=o, mark options={solid}]
table[x=para, y=gowalla_hr] {latFactor.txt};
\addplot[color={rgb:red,0;green,157;blue,178}, mark=square, mark options={solid}]
table[x=para, y=ml10m_hr] {latFactor.txt};
\addplot[color={rgb:red,245;green,9;blue,11}, mark=triangle, mark options={solid}]
table[x=para, y=amazon_hr] {latFactor.txt};
\addplot[color={rgb:red,10;green,10;blue,240}, mark=x, mark options={solid}]
table[x=para, y=yelp_hr] {latFactor.txt};
\legend{\large Gowalla, \large MovieLens, \large Amazon, \large Yelp};
\end{axis}
\begin{axis}[
width=12cm, 
    height=9cm, 
xmin=4,xmax=14,
ymin=-40,ymax=5,
xtick distance=2, 
axis x line=top,
axis y line=none,
xlabel={Number of Short-term Graphs $T$ for Yelp},
ylabel={},
label style={scale=2.2},
every tick label/.append style={scale=1.8},
]
\end{axis}
\end{tikzpicture}

\begin{tikzpicture}
\begin{axis}[
    width=12cm, 
    height=9cm, 
    xlabel={User Personalized Weight $d_{sal}$},
    ylabel={Decrease of HR@10 (\%)},
    xmin=16,xmax=48,
    ymin=-15,ymax=1,
    legend columns=1,
    legend cell align=right,
    grid=both,
    every axis plot/.append style={ultra thick},
    every tick label/.append style={scale=1.7},
    label style={scale=2.5},
    legend style={
        nodes={scale=1.5, transform shape},
        legend image post style={scale=1.5},
        },
    legend style={at={(1,0)},anchor=south east},
    every axis plot post/.append style={
        every mark/.append style={scale=2}
    },
]
\addplot[color={rgb:red,133;green,76;blue,255}, mark=o, mark options={solid}]
table[x=para, y=gowalla_hr] {hyperNum.txt};
\addplot[color={rgb:red,0;green,157;blue,178}, mark=square, mark options={solid}]
table[x=para, y=ml10m_hr] {hyperNum.txt};
\addplot[color={rgb:red,245;green,9;blue,11}, mark=triangle, mark options={solid}]
table[x=para, y=amazon_hr] {hyperNum.txt};
\addplot[color={rgb:red,10;green,10;blue,240}, mark=x, mark options={solid}]
table[x=para, y=yelp_hr] {hyperNum.txt};
\legend{\large Gowalla, \large MovieLens, \large Amazon, \large Yelp};
\end{axis}
\end{tikzpicture}

\begin{tikzpicture}
\begin{axis}[
    width=12cm, 
    height=9cm, 
    xlabel={Weight for SAL Loss $\lambda_1$ ($1e^{-x}$)},
    ylabel={Decrease of HR@10 (\%)},
    xmin=4,xmax=7,
    ymin=-30,ymax=1,
    legend columns=1,
    legend cell align=right,
    grid=both,
    every axis plot/.append style={ultra thick},
    label style={scale=2.5},
    every tick label/.append style={scale=1.7},
    legend style={
        nodes={scale=1.5, transform shape},
        legend image post style={scale=1.5},
        },
    legend style={at={(1,0)},anchor=south east},
    every axis plot post/.append style={
        every mark/.append style={scale=2}
    }
]
\addplot[color={rgb:red,133;green,76;blue,255}, mark=o, mark options={solid}]
table[x=para, y=gowalla_hr] {lamda.txt};
\addplot[color={rgb:red,0;green,157;blue,178}, mark=square, mark options={solid}]
table[x=para, y=ml10m_hr] {lamda.txt};
\addplot[color={rgb:red,245;green,9;blue,11}, mark=triangle, mark options={solid}]
table[x=para, y=amazon_hr] {lamda.txt};
\addplot[color={rgb:red,10;green,10;blue,240}, mark=x, mark options={solid}]
table[x=para, y=yelp_hr] {lamda.txt};
\legend{\large Gowalla, \large MovieLens, \large Amazon, \large Yelp};
\end{axis}
\end{tikzpicture}

%% file: conclusion.tex
\section{Conclusion}
\label{sec:conclusoin}
In this paper, we explore sequential recommender systems with graph neural networks and introduce a novel personalized self-augmented learning paradigm to boost recommendation robustness. Our \model\ framework refines interest representation by integrating collaborative patterns and user behavior sequences, employing self-augmented learning to adaptively reduce short-term noise based on user-specific stability traits. Extensive experiments confirm the model's enhanced performance and denoising ability over existing baselines. For future work, we aim to investigate adaptive dynamic short-term graph partitioning techniques to more accurately capture short-term characteristics across various datasets, further enhancing the recommendation performance.
